\documentclass{article}

\usepackage{PRIMEarxiv}

\usepackage[utf8]{inputenc} 
\usepackage[T1]{fontenc}    
\usepackage{hyperref}       
\usepackage{url}            
\usepackage{booktabs}       
\usepackage{amsfonts}       
\usepackage{nicefrac}       
\usepackage{microtype}      
\usepackage{lipsum}
\usepackage{fancyhdr}       
\usepackage{graphicx}       
\usepackage{amsmath}
\usepackage{algorithm}
\usepackage{algorithmic}
\usepackage{multirow} 
\graphicspath{{media/}}     

\title{KAN-HAR: A Human activity recognition based on Kolmogorov-Arnold Network
}

\author{
  Mohammad Alikhani \\
  Faculty of Electrical Engineering \\
  K.N. Toosi University of Technology \\
  Tehran, Iran\\
  \texttt{m.alikhani2@email.kntu.ac.ir}
}

\begin{document}
\maketitle

\begin{abstract}
Human Activity Recognition (HAR) plays a critical role in numerous applications, including healthcare monitoring, fitness tracking, and smart environments. Traditional deep learning (DL) approaches, while effective, often require extensive parameter tuning and may lack interpretability. In this work, we investigate the use of a single three-axis accelerometer and the Kolmogorov–Arnold Network (KAN) for HAR tasks, leveraging its ability to model complex nonlinear relationships with improved interpretability and parameter efficiency. The MotionSense dataset, containing smartphone-based motion sensor signals across various physical activities, is employed to evaluate the proposed approach. Our methodology involves preprocessing and normalization of accelerometer and gyroscope data, followed by KAN-based feature learning and classification. Experimental results demonstrate that the KAN achieves competitive or superior classification performance compared to conventional deep neural networks, while maintaining a significantly reduced parameter count. This highlights the potential of KAN architectures as an efficient and interpretable alternative for real-world HAR systems.
The open-source implementation of the proposed framework is available at the Project's \href{https://github.com/BahMoh/KAN-HAR}{GitHub Repository}. 
\end{abstract}


\section{Introduction}\label{sec_Introduction}
Machine learning (ML) and DL have become pivotal technologies in modern data-driven systems, enabling computers to learn from data and make decisions without explicit programming. Over the past decade, DL models, in particular, have demonstrated remarkable capabilities in extracting high-level abstractions from raw data, achieving state-of-the-art performance in a wide range of domains. These include computer vision tasks such as image classification and object detection, natural language processing for machine translation and sentiment analysis, and speech recognition in voice assistants. Beyond these, ML and DL have also transformed healthcare through medical image analysis, disease prediction \cite{yu2023popular}, and personalized treatment \cite{wen2023deep}, revolutionized autonomous systems in robotics and self-driving vehicles \cite{alikhani2026learning, alikhani2025long}, and powered intelligent applications in finance \cite{gadre2016review}, cyber-security \cite{alikhani2025contrastive, alikhani2026dkd}, and industrial automation \cite{iqbal2019fault}. This widespread adoption is driven by their ability to model complex, nonlinear relationships in data, making them indispensable tools in both academic research and real-world applications.

The rapid growth of the IoT has further expanded the scope of intelligent systems by enabling large-scale, continuous data collection from interconnected devices. With billions of sensors embedded in smartphones, wearable devices, and smart environments, IoT systems generate rich streams of multi-modal data that can be leveraged for context-aware and personalized services. Among the many IoT-driven applications, HAR \cite{kumar2024human} has emerged as a key research area due to its potential in enhancing healthcare monitoring, elderly care, fitness tracking, workplace safety, and human–computer interaction. HAR systems aim to identify and classify physical activities, such as walking, running, sitting, or climbing stairs, based on motion and physiological signals. In recent years, advances in embedded sensing technologies, particularly low-cost inertial measurement IMUs in consumer devices \cite{kim2020machine}, have enabled accurate activity recognition in real-world settings without the need for intrusive or expensive equipment. Despite these advances, HAR still faces several real-world challenges, including inter-person variability in movement patterns across different users, environmental noise caused by background vibrations or variations in device placement, resource constraints related to battery life, computational power, and memory in IoT devices, and scalability concerns when deploying HAR systems to millions of devices in diverse environments. In this work, we address these challenges with a particular focus on mitigating inter-person variability, where our proposed approach demonstrates robustness to differences in individual motion patterns while maintaining high classification accuracy.

The contributions of this study are as follows:

\begin{itemize}
    \item Transparent and interpretable modeling, utilizing the Kolmogorov-Arnold representation theorem
    \item Open-source implementation available on the GitHub repository of the project
    \item High-performing KAN model for human activity recognition 
    \item The challenge of inter-person variability is addressed by ensuring that users in the training set are separate from those in the test set, improving the model's ability to generalize across different individuals.
\end{itemize}

The rest of the paper is organized as follows:
\autoref{sec:Related Works} reviews the related works in the field.
\autoref{sec_Datasets} describes the datasets used in this study, with details of the MotionSense dataset provided in \autoref{subsec_motionsense}.
\autoref{sec:Method} outlines our proposed methodology, including the Kolmogorov–Arnold Network (KAN) in \autoref{subsec_KAN}, the preprocessing pipeline in \autoref{subsec_preprocessing}, and the experimental setup in \autoref{subsec:setup}.
The experimental results are presented and discussed in \autoref{sec_results}.
Finally, \autoref{sec_Conclusion} concludes the paper and outlines potential directions for future work.

\section{Related Works}\label{sec:Related Works}
The study in \cite{xu2019innohar} addresses the limitations of traditional ML approaches, which have typically depended on handcrafted features and statistical models, often proving inadequate for highly complex and irregular waveform data. This work leverages DL by integrating inception-based convolutional modules with gated recurrent units (GRUs) to capture multi-dimensional features and model temporal dependencies directly from raw multi-channel sensor inputs.

The study in \cite{khan2022human} tackles the difficulties in capturing both spatial and temporal features in HAR, as well as the limitations of existing datasets. To overcome these issues, the authors propose a hybrid CNN-LSTM model, where CNNs are employed for spatial feature extraction and LSTMs for modeling temporal dependencies, evaluated on a newly collected Kinect V2 dataset comprising 12 different human activity classes.

In \cite{murad2017deep} a LSTM network is proposed for human activity recognition, capable of capturing long-range temporal dependencies in variable-length input sequences from body-worn sensors, overcoming the limitations of conventional ML and CNN-based methods. 
\cite{abdel2020st} proposes a supervised dual-channel model combining LSTM with an attention mechanism for temporal fusion and a convolutional residual network for spatial fusion, enhanced by an adaptive channel-squeezing operation to exploit multichannel dependencies.

In \cite{hassan2018human} frames HAR as a classification problem using a Deep Belief Network (DBN), with features preprocessed via KPCA and LDA to improve robustness.
\cite{zhu2018novel} proposes a semi-supervised DL approach combining LSTM-based temporal ensembling with both labeled and unlabeled data, enabling the model to capture temporal dependencies and leverage unlabeled data effectively.
\cite{ronao2016human} proposes a deep 1D convnet to automatically extract robust features from raw smartphone sensor data.

\cite{ravi2016deep} proposes a DL-based HAR technique for real-time classification on low-power wearable devices, using spectral-domain feature generation to achieve invariance to sensor orientation, placement, and acquisition rates. 
Authors in \cite{jaouedi2020new} propose a hybrid DL model leveraging gated recurrent neural networks (GRNNs) for sequential data, emphasizing robust feature extraction to improve classification performance. 

\cite{gupta2021deep} proposes a hybrid CNN-GRU model for human activity recognition, where CNNs capture spatial features from raw inertial sensor data and GRUs model temporal dependencies. The approach automatically extracts robust features without manual engineering.
A CNN-based feature learning method is proposed in \cite{wang2018spatial} that automatically extracts features from raw inputs. The effect of hyper-parameters like layer number and kernel size extensively analyzed for performance

\cite{gumaei2019hybrid} proposes a hybrid DL model combining simple recurrent units (SRUs) and GRUs to effectively process sequences, manage temporal dependencies, and mitigate vanishing gradient issues. 
The study in \cite{moshiri2021csi} leverages WiFi Channel State Information (CSI) for HAR, offering a non-intrusive alternative to vision and sensor-based methods. CSI data were collected using a Raspberry computer, converted to images, and classified with a 2D CNN. 
The study in \cite{abbaspour2020comparative} investigates hybrid DL models that combine CNNs with various RNNs (LSTM, BiLSTM, GRU, BiGRU) to capture both spatial and temporal features.

In the majority of recent HAR studies, DL models are employed to automatically extract features directly from raw sensor data. While this approach benefits from representation learning and can uncover complex patterns without manual intervention, it also presents several limitations. First, deep models often require large amounts of labeled data to generalize well, which may not be available in many real-world HAR applications. Second, they can be computationally expensive and difficult to deploy on low-power wearable devices or IoT systems. Third, automatically learned features may be less interpretable, making it challenging to understand what patterns the model relies on for classification. Finally, DL models can be sensitive to noise, sensor placement, and inter-person variability, leading to inconsistent performance across different users or environments.

In contrast, handcrafted features, carefully designed based on domain knowledge, allow for compact, interpretable, and robust representations that can perform well even with limited data and constrained computing resources. Therefore, in this work, we adopt handcrafted features to mitigate these challenges while maintaining effective performance in real-world HAR scenarios.

\section{Datasets} \label{sec_Datasets}
In the field of HAR, there exist various types of datasets that differ in their sensor modalities, number of participants, sampling frequencies, and types of activities performed. Some datasets are collected using smartphones, such as UCI-HAR \cite{anguita2013public} and WISDM \cite{weiss2019wisdm}, while others use wearable devices such as smartwatches or multiple body-worn sensors, for example, PAMAP2 \cite{reiss2012introducing}. Some datasets include multi-modal recordings from several sensor types simultaneously, such as the Opportunity dataset \cite{chavarriaga2013opportunity}. These datasets provide valuable benchmarks for evaluating the performance and generalization capabilities of HAR algorithms across different scenarios and environments. In this work, we evaluate the performance of the proposed method on the MotionSense dataset.

\subsection{MotionSense Dataset}\label{subsec_motionsense}
The MotionSense dataset is a publicly available collection of smartphone sensor data designed for HAR and demographic attribute analysis. Data were collected from 24 participants using an iPhone 6s placed in the front pocket, with measurements recorded via the accelerometer and gyroscope at a sampling rate of 50 Hz using the SensingKit framework. Six activities were performed, including walking, jogging, sitting, standing, upstairs, and downstairs. Each activity was repeated across 15 trials to ensure variability \cite{Malekzadeh:2018:PSD:3195258.3195260, Malekzadeh:2019:MSD:3302505.3310068}. The dataset contains 12 channels, including 3-axis accelerometer signals (attitude, gravity, user acceleration) and 3-axis gyroscope signals (rotation rate). In this work, we use only the 3-axis accelerometer signals, from which we extract 12 features per channel, resulting in a total of 36 features. These features are: mean absolute value, standard deviation, skewness, kurtosis, Shannon entropy, root mean square (RMS), maximum absolute value, peak-to-peak (P2P), crest factor, clearance factor, shape factor, and impulse factor. \autoref{tab:time_features_formulas} presents the extracted time-domain features per accelerometer axis along with their corresponding formulas.

\begin{table}[h]
\centering
\caption{Extracted Time-Domain Features per Accelerometer Axis with Formulas}
\label{tab:time_features_formulas}
\renewcommand{\arraystretch}{2.5} 
\begin{tabular}{@{}lll@{}}
\toprule
\textbf{No.} & \textbf{Feature} & \textbf{Formula} \\ \midrule
1 & Mean Absolute Value & $\displaystyle \text{MAV} = \frac{1}{N} \sum_{i=1}^{N} |x_i|$ \\
2 & Standard Deviation & $\displaystyle \sigma = \sqrt{\frac{1}{N} \sum_{i=1}^{N} (x_i - \bar{x})^2}$ \\
3 & Skewness & $\displaystyle \text{Skew} = \frac{1}{N} \sum_{i=1}^{N} \left( \frac{x_i - \bar{x}}{\sigma} \right)^3$ \\
4 & Kurtosis & $\displaystyle \text{Kurt} = \frac{1}{N} \sum_{i=1}^{N} \left( \frac{x_i - \bar{x}}{\sigma} \right)^4 - 3$ \\
5 & Shannon Entropy & $\displaystyle H = -\sum_{j} p_j \log p_j$ \\
6 & Root Mean Square (RMS) & $\displaystyle \text{RMS} = \sqrt{\frac{1}{N} \sum_{i=1}^{N} x_i^2}$ \\
7 & Maximum Absolute Value & $\displaystyle x_\text{max} = \max |x_i|$ \\
8 & Peak-to-Peak (P2P) & $\displaystyle \text{P2P} = \max(x_i) - \min(x_i)$ \\
9 & Crest Factor & $\displaystyle \text{Crest} = \frac{x_\text{max}}{\text{RMS}}$ \\
10 & Clearance Factor & $\displaystyle \text{Clearance} = \left( \frac{1}{N} \sum_{i=1}^{N} \sqrt{|x_i|} \right)^2$ \\
11 & Shape Factor & $\displaystyle \text{Shape} = \frac{\text{RMS}}{\text{MAV}}$ \\
12 & Impulse Factor & $\displaystyle \text{Impulse} = \frac{x_\text{max}}{\text{MAV}}$ \\ \bottomrule
\end{tabular}
\end{table}

\section{Methodology}\label{sec:Method}
This section describes the methodology proposed in this work. In Section \autoref{subsec_KAN}, we first provide an overview of the Kolmogorov–Arnold representation theorem and the KAN model. Then, \autoref{subsec_preprocessing} details the pre-processing procedures applied in the proposed approach.

\subsection{KAN: Kolmogorov-Arnold Network}\label{subsec_KAN}
Introduced in \cite{liu2024kan}, the Kolmogorov–Arnold Network (KAN) serves as an alternative to conventional fully connected networks, offering strong capabilities in modeling non-linear functions and outperforming traditional MLP layers. KAN has demonstrated substantial potential across various domains, including anomaly detection \cite{abudurexiti2025explainable}, human activity recognition \cite{liu2024ikan}, differential equation solving \cite{ma2024integrating}, and time series analysis \cite{lee2024hippo}. Noteworthy achievements include improvements in accuracy, interpretability, and computational efficiency \cite{liu2024kan}.

Before introducing the proposed methodology, we review existing applications of KANs in the field of HAR. In \cite{liu2024initial} the possibility of using KAN as a feature extractor for IMU-base HAR is investigated, leveraging their spline-based non-linear computations to learn complex functions more efficiently than conventional networks. The work in \cite{costa2024real} evaluates KANs on real-world classification tasks, enhances HAR, and explores real-time HAR on mobile devices. Experiments with seven kernels show that KANs outperform traditional ML and MLP models while maintaining low latency.

Authors in \cite{liu2024ikan} proposed iKAN framework introduces incremental learning for wearable sensor HAR by replacing MLP classifiers with KAN to exploit spline-based local plasticity and global stability. It expands task-specific feature extraction branches to handle new sensor modalities without altering classifier dimensions, achieving superior continual learning performance across six public HAR datasets. \cite{khan2024kgan} proposes a KGAN framework that combines the Kolmogorov–Arnold representation with a GAN to enable semi-supervised domain adaptation for HAR, generating synthetic target-domain data to address label scarcity and improve robustness. It integrates kernel mean matching (KMM) with maximum mean and covariance discrepancy (MMCD) to better align source and target distributions.

KANs are grounded in the Kolmogorov–Arnold representation theorem, which was originally formulated to solve differential equations. In this study, however, KAN is applied to classification tasks. The theorem guarantees that any multivariate continuous function $f(X)$, where $X = (X_1, \dots, X_T)$, can be decomposed into a sum of nested univariate functions. Specifically, for a smooth differentiable function $f: [0,1]^T \rightarrow \mathbb{R}^{d_{\text{out}}}$, there exist continuous univariate functions $\Phi_q$ and $\phi_{q,p}$ that satisfies:
\begin{equation}
f(X) = \sum^{2T+1}_{q=1}\Phi_q \bigl ( \sum^T_{p=1}\phi_{q,p}(X_p) \bigr )
\label{eq:KA_theorem}
\end{equation}
where $\phi_{q,p}: [0, 1] \rightarrow \mathbb{R}$ and $\Phi_q: \mathbb{R} \rightarrow \mathbb{R}$. In this formulation, each $\phi_{q,p}$ and $\Phi_q$ is a univariate function, and the nested sum of these functions enables the approximation of any continuous multivariate function. The Kolmogorov–Arnold representation theorem guarantees that $(2T+1) \times T$ univariate functions are sufficient to approximate any given $T$-variate function. 

Compared to Taylor or Fourier series approximations, which may require an infinite number of terms to represent a multivariate function with high precision, KANs offer a significant advantage by approximating a function using a finite composition of learnable univariate functions.

\begin{figure}
    \centering
    \includegraphics[width=0.5\linewidth]{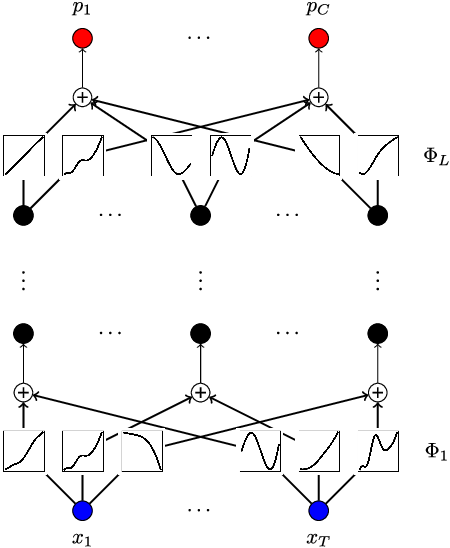}
    \caption{Visualization of a stack of $L$ KAN layers with an input dimension of $T$ and an output dimension of $C$, where $C$ denotes the number of classes \cite{baravsin2024exploring}.}
    \label{fig_KAN_architecture}
\end{figure}

KANs leverage this theoretical foundation by replacing the typical pre-defined activation functions in neural networks with learned univariate functions applied to each input dimension individually. Unlike traditional MLP layers, where activation functions are applied to the outputs of nodes, KAN places activation functions on the edges of the network, i.e., between connected nodes. \autoref{fig_KAN_architecture} illustrates a KAN with an input size of $T$ and an output size of $C$. 

From this point on, to model the approximated function $f(\cdot)$, we use the notation $KAN(\cdot)$. Here, $X^i \in \mathbb{R}^T$ denotes a single sample with a corresponding label $Y^i \in \{1, \dots, C\}$. In \autoref{fig_KAN_architecture}, the output has dimension $C$. More formally, we can write $p = KAN(X^i)$, where $p = [p_1, p_2, \dots, p_C]$ is a vector representing the class scores.

The forward propagation of a KAN model with $L$ layers can be expressed as a composition of non-linear transformations $\Phi_q$, each applied to the output of the previous layer:
\begin{equation}
y = KAN(X) = (\Phi_L \circ \Phi_{L-1} \circ \dots \circ \Phi_{1})X
\label{eq:kan_in_matrix_form}
\end{equation}
each $\Phi_l$ consists of a set of learnable univariate functions $\phi$ applied element-wise. In contrast to traditional MLPs, where nonlinearity is introduced via fixed activation functions, KANs learn the activation functions during training. A traditional MLP network performs forward computation using linear weight matrices followed by fixed nonlinear activations:
\begin{equation}
y = MLP(X) = (W_L \circ \sigma \circ W_{L-1} \circ \sigma \circ \dots \circ W_{1})X
\label{eq_mlp_in_matrix_form}
\end{equation}
here, $W_l$ are the weight matrices and $\sigma$ denotes a fixed activation function that introduces nonlinearity into the output of each layer. In contrast, KANs learn the activation functions instead of using pre-defined ones. KANs employ a hybrid activation function $\phi(X)$, which combines the SiLU and spline-based basis functions. This function $\phi(X)$, is defined as follows:
\begin{equation}\label{eq:silu_spline}
\phi(X) = \omega_bSiLU(X)+\omega_sSpline(X)
\end{equation}
where the SiLU activation function will be defined by:
\begin{equation}\label{eq:silu_activation}
SiLU(X) = \frac{X}{1 + e^{-X}}
\end{equation}

The SiLU activation function is a continuous and differentiable function that ensures gradient flow during training, providing favorable optimization properties. The B-splines enable fine-grained modeling of nonlinearities through a learnable basis function. Each spline function operates within a grid of size $G$:
\begin{equation}\label{eq:spline_basis_function}
Spline(X) = \sum ^{G+k}_{i=1} c_iB_i(X)
\end{equation}
here, $B_i(X)$ denotes a spline basis function, $c_i$ represents a learned coefficient or control point, $G$ is the number of grid points, and $K$ denotes the spline order (by default, $K=3$ for cubic splines). The spline order $K$ determines the smoothness of the curve, while a larger grid size $G$ increases the resolution of the splines. As shown in \autoref{eq:spline_basis_function}, a spline with order $K$ and grid size $G$ requires $G+K$ basis functions. The total number of learnable parameters in a KAN layer is calculated as follows:

\begin{equation}\label{eq:KAN_parameters}
Parameters = (d_{in} \times d_{out})(G+K+3) + d_{out}
\end{equation}
here, $d_{\text{in}}$ and $d_{\text{out}}$ are the input and output dimensions, $G$ is the grid size (number of grid points), and $K$ is the spline order. Our implementation employs two stacked KAN layers: the first layer serves as a feature extractor, and the second performs the final classification.

Unlike traditional MLPs, where the number of parameters depends solely on the input and output dimensions, in KAN the spline order $K$ and grid size $G$ also influence the total parameter count. As a result, KAN may have more parameters than an MLP with equivalent input and output dimensions.

\subsection{Pre-processing}\label{subsec_preprocessing}
In this study, the only pre-processing applied is standardization using the data’s mean ($\mu$) and standard deviation ($\sigma$). KAN’s inherent capability allows it to classify intrusion data without further pre-processing, enabling faster inference. The standardization procedure is defined as follows:
\begin{equation}
X_{\text{scaled}} = \frac{X - \mu}{\sigma}
\label{eq_standard_scaler}
\end{equation}

\subsection{Simulation Setup} \label{subsec:setup}
The implementation of this work is available at GitHub Repository\footnote{\url{https://github.com/BahMoh/KAN-HAR}}. Adam is used as the optimizer during pre-training, whereas AdamW is employed during fine-tuning. All experiments are conducted in the Kaggle environment, which is equipped with 32 GB of RAM, an NVIDIA Tesla P100 GPU with 16 GB of VRAM, and an Intel(R) Xeon(R) CPU running at 2.00 GHz. 
The dataset comprises recordings from 24 participants, with no overlap between individuals in the training and test sets. We use data from 19 participants for training and from 5 participants for testing.

\section{Results}\label{sec_results}
\begin{figure}
    \centering
    \includegraphics[width=0.7\linewidth]{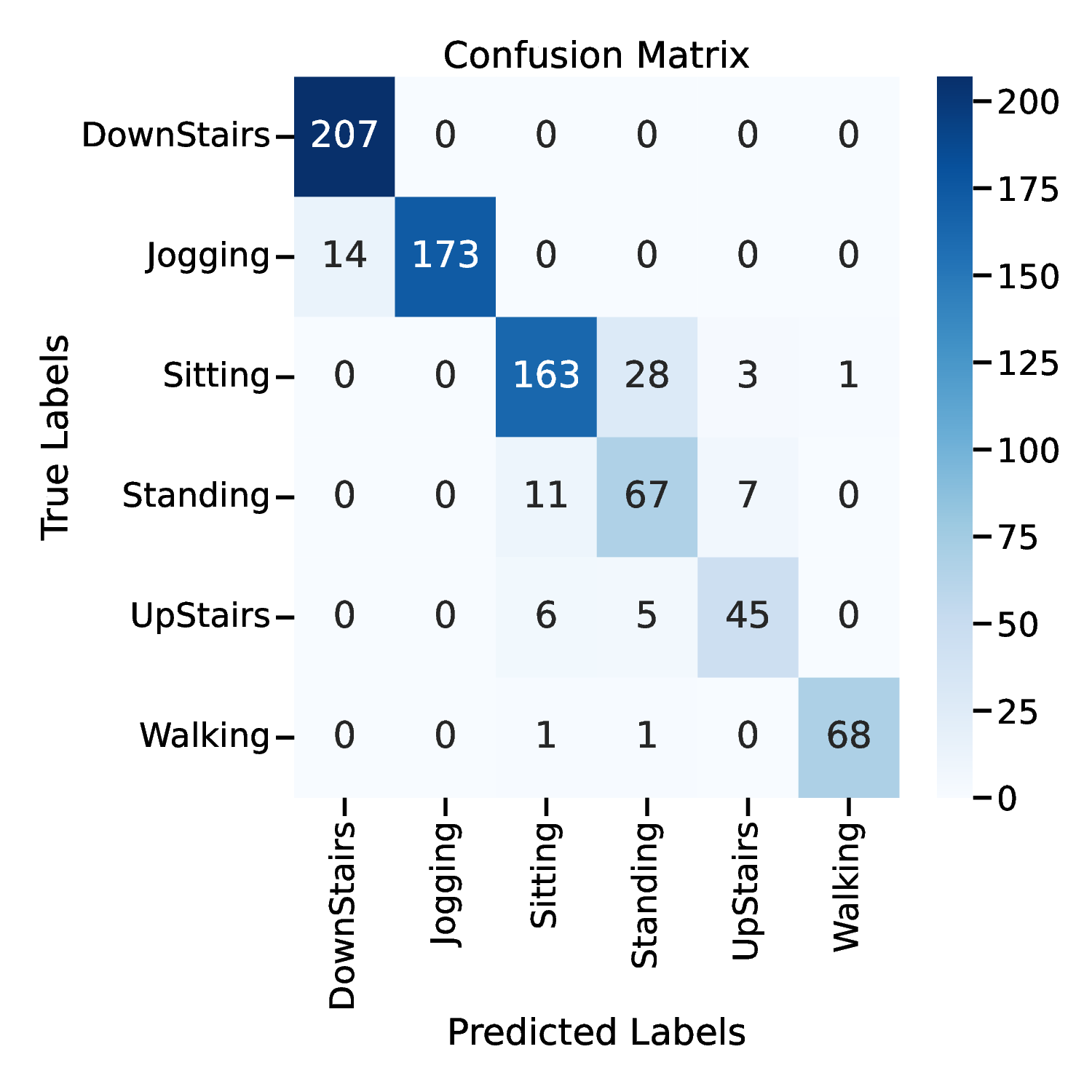}
    \caption{Confusion matrix of the MotionSense dataset}
    \label{conf_matrix}
\end{figure}

\autoref{conf_matrix} presents the confusion matrix of the experimental results, demonstrating a high level of classification accuracy with minimal misclassifications, indicating the effectiveness of the proposed approach.

\autoref{tab_motionsense_comparison} presents a performance comparison between the proposed method and existing approaches from the literature. The proposed KAN model achieves the highest overall accuracy and F1-score, demonstrating strong inter-person generalization capability and competitive precision and recall, thereby validating its effectiveness for cross-participant HAR scenarios.

\begin{table*}[]
\caption{Performance comparison of the proposed HAR method with other appearances in the literature}
\label{tab_motionsense_comparison}
\centering
\begin{tabular}{llllll}
\toprule
\textbf{Literature} & \textbf{Method} & \textbf{Accuracy (\%)} & \textbf{Precision (\%)} & \textbf{Recall (\%)}  & \textbf{F1 (\%)} \\ \midrule
\cite{batool2019sensors}              &  MFCC + SVM
                          & 85.35                      & 85.93               & 85.68          & 85.80             \\
\cite{jalal2020stochastic}            &  DT + BGWO
                          & 87.71                      & 87.21               & 87.21          & 86.93             \\
\cite{saeed2019multi}                &  Self-supervised TPN
                          & 89.01                      & 89.01               & 88.99          & 88.99             \\
\cite{saha2024decoding}            &  FusionActNet   
                          & 90.35                      & \textbf{90.00 }     & \textbf{90.07} & 90.21                 \\ \midrule
Proposed Method & KAN                 
                          & \textbf{90.38}             & 87.75               & 86.83          & \textbf{90.52} \\ \bottomrule
\end{tabular}
\end{table*}

\section{Conclusion} \label{sec_Conclusion}
In this study, we employed the KAN for HAR using handcrafted features. The proposed method demonstrated strong classification performance, as evidenced by the experimental results, and outperformed existing approaches in comparative evaluations. The reliance on handcrafted features, combined with the expressive capability of KAN, allowed our model to achieve competitive accuracy and generalization, particularly in cross-participant scenarios. The implementation is publicly available on our \href{https://github.com/BahMoh/KAN-HAR}{GitHub Repository}, facilitating reproducibility and future research extensions.
\bibliographystyle{unsrt}  
\bibliography{references}

\end{document}